# A Predictive Structural Model of the Primate Connectome

**Sarah F. Beul[1], Helen Barbas[2,3], Claus C. Hilgetag[1,2*]**

[1] Department of Computational Neuroscience, University Medical Center Hamburg-Eppendorf, Martinistr.52 – W36, 20246 Hamburg, Germany

[2] Neural Systems Laboratory, Department of Health Sciences, Boston University, Commonwealth Ave. 635, 20115 Boston, MA, USA

[3] Boston University School of Medicine, Department of Anatomy and Neurobiology, 72 East Concord St., 02118 Boston, MA, USA

**Corresponding author:** Claus C. Hilgetag, Department of Computational Neuroscience, University Medical Center Hamburg-Eppendorf, Martinistr.52 – W36, 20246 Hamburg, Germany, c.hilgetag@uke.de

Abstract

Anatomical connectivity imposes strong constraints on brain function, but there is no general agreement about principles that govern its organization. Based on extensive quantitative data we tested the power of three models to predict connections of the primate cerebral cortex: architectonic similarity (structural model), spatial proximity (distance model) and thickness similarity (thickness model). Architectonic similarity showed the strongest and most consistent influence on connection features. This parameter was strongly associated with the presence or absence of inter-areal connections and when integrated with spatial distance, the model allowed predicting the existence of projections with very high accuracy. Moreover, architectonic similarity was strongly related to the laminar pattern of projection origins, and the absolute number of cortical connections of an area. By contrast, cortical thickness similarity and distance were not systematically related to connection features. These findings suggest that cortical architecture provides a general organizing principle for connections in the primate brain.

# Introduction

Structural connections impose strong constraints on functional interactions among brain areas [1]. It is thus essential to understand the principles that underlie the organization of connections which give rise to the topological properties of the cortex.

Global brain connectivity is neither random nor regular. Moreover, there are striking regularities in the laminar patterns of projection origins and terminations [2–5]. Large-scale topological features of brain networks include modules and highly connected hubs [6]. Other prominent topological features are hub-modules, so-called 'rich-clubs' or 'network cores', which have been identified in structural and functional neural networks of several species [7].

The presence of nonrandom features in brain networks points to the existence of organizing factors. We hypothesize that inherent structural properties of the cortex account for prominent characteristics of the cortical connectome. Here, we investigated to which extent three principal structural factors explain connection features.

The first factor is cortical architecture, which has been used to formulate a relational 'structural model' [8,9]. The model relies on the relative architectonic similarity between linked areas to predict the laminar distribution of their interconnections. The structural model is based on evidence that architectonic features change systematically within cortical systems [10]; reviewed in [11,12]. Cortical architecture can be defined by a number of structural features, including the neuronal density of cortical areas, as well as the number of identifiable cortical layers, myelin density and a number of receptor markers and specialized inhibitory neurons [13–17]. By capitalizing on cortical architecture, the structural model explains the laminar origin and termination patterns of ipsilateral and contralateral corticocortical connections in the macaque prefrontal and cat visual cortex [8,9,18,19], as well as existence of projections and topological properties of individual areas across the entire cat cortex [20].

As a second feature we considered the spatial proximity of cortical areas. In the 'distance model', the spatial separation of areas is hypothesized to account for the existence [21–23], strength [24,25] as well as laminar patterns [26] of corticocortical projections. According to the distance model, connections between remote areas are less frequent and sparser than connections among close areas.

One other factor that has received much attention in the study of possible relations between brain morphology and connectivity is cortical thickness, an attractive possibility, because thickness can be assessed non-invasively by MRI. Cortical thickness has been related to neuron density [27,28] and suggested as an indicator of overall cortical composition [29–31]. Cortical thickness covariations have been treated as a surrogate of anatomical connectivity (but see [32]. The inferred structural networks based on cortical thickness have been explored with respect to their topological properties, association with functional connectivity, and relationship to behavioral traits (e.g., [33–36]; for a review see [37]. Given this strong interest in the possible significance of cortical thickness, we assessed this parameter as an anatomical covariate of structural connectivity ('thickness model').

We compared the predictive power of the three models for connection data from a comprehensive data set (*connectome*). The tested database provides extensive quantitative information on the existence and laminar origins of projections linking cortical areas in the macaque brain [38,39]. We investigated whether this connectome can be understood in terms of the underlying brain anatomy.

## Results

We examined the association between the primate cortical connectome and these anatomical features of the primate cerebral cortex: neuron density (a quantitative measure of cortical architecture, Fig 1); spatial proximity; and cortical thickness. We tested how well each of the three anatomical parameters was related to the existence and the laminar origins of projections between cortical areas, and could predict the presence or absence of projections. We found that the existence of projections is most closely related to the neuron density of cortical areas. We also showed that neuronal density is the anatomical factor that best explains laminar projection patterns and is linked to topological properties of brain regions.

**Relations among anatomical variables**

To quantify relative structural similarity across the cortex, for all pairs of connected areas we computed the difference in neuron density or cortical thicknesses as measured on a log scale. That is, structural (dis-)similarities were expressed as log-ratios. Spatial proximity was quantified by Euclidean distance between areas. The anatomical variables associated with the corticocortical projections were not completely independent. We found a moderate correlation between the undirected neuron density ratio and the Euclidean distance of area pairs ($r = 0.47$, $p < .001$), whereas the correlation of Euclidean distance with the undirected thickness ratio was significant but of negligible size ($r = 0.12$, $p < .001$). In contrast, neuron density ratio and thickness ratio were strongly negatively correlated ($r = -0.76$, $p < .001$), an association which results from a strong inverse correlation between the neuron density and thickness of brain areas ($r = -0.69$, $p < .001$).

**Existence of projections**

We used three different approaches to explore how the three anatomical variables of cortical density, thickness and distance relate to the absence and presence of projections. In an initial comparison, we found that connected areas were closer or more similar than non-connected areas, for all three structural parameters (mean |log-ratio$_{density}$|(absent) = 0.49, mean |log-ratio$_{density}$|(present) = 0.24, $t_{(1126)} = 13.8$, $p < .001$; mean distance(absent) = 32.9mm, mean distance(present) = 25.7mm, $t_{(2608)} = 15.1$, $p < .001$;

mean |log-ratio$_{thickness}$|(absent) = 0.20, mean |log-ratio$_{thickness}$|(present) = 0.14, $t_{(2608)}$ = 11.5, $p < .001$). This effect was largest for the neuron density ratio (effect sizes: |log-ratio$_{density}$|: $r = 0.38$, distance: $r = 0.28$, |log-ratio$_{thickness}$|: $r = 0.22$).

Then, to assess the distribution of absent and present projections across the three structural variables in more detail, we plotted the relative frequency of present projections across neuron density ratio and Euclidean distance in comparison to the absolute numbers of absent and present projections (Fig 2). For all variables, present projections became relatively less frequent with increasing distance or structural dissimilarity of two potentially connected areas, as also shown by a rank correlation, ρ, of the relative frequencies (|log-ratio$_{density}$|: ρ = -1.00, $p < .001$; distance: ρ = -0.98, $p < .001$; |log-ratio$_{thickness}$|: ρ = -0.93, $p < .01$).

Finally, to exploit the association of the structural variables with the existence of cortical connections, we used the parameters to classify projections as either absent or present. We predicted projection presence or absence based on all 7 possible combinations of the three parameters (each individual parameter, 3 pairwise combinations of the parameters, and a combination of all three parameters). Ssee Methods for a detailed description of the classification and validation procedure.

The best classification among the 6 combinations of one or two parameters was obtained from the combination of the log-ratio of neuron density (i.e., density similarity) with Euclidean distance. This pairing was superior to all other combinations; its accuracy, precision and negative predictive value were not exceeded at comparable thresholds, and overall performance as quantified by the mean Youden-index *J* was worse for all other combinations (mean ± standard deviation:*J*(|log-ratio$_{density}$| & distance) = 0.75 ± 0.04; *J*(distance & |log-ratio$_{thickness}$|) = 0.51 ± 0.13; *J*(|log-ratio$_{density}$| & |log-ratio$_{thickness}$|) = 0.11 ± 0.03; *J*(|log-ratio$_{density}$|) = 0.0 ± 0.0; *J*(distance) = 0.07 ± 0.03; *J*(|log-ratio$_{thickness}$|): no predictions at thresholds above *p(present)* = 0.775; see Supplementary Figure S1 for the underlying distribution of true positive rate and false positive rate and Supplementary Figure S2 for a detailed depiction of the Youden-index $J$ across all thresholds). Including all three anatomical variables as predictive variables did not improve classification accuracy or overall performance as assessed by the mean Youden-index ($J$(|log-ratio$_{density}$| & distance & |log-ratio$_{thickness}$|) = 0.76 ± 0.04 (Fig. 3C). A Kruskal-Wallis-test showed that the distributions of the Youden-index $J$ were significantly different between the combinations of the parameters ($H$ = 549.2, $p$ < .001). *Post hoc* tests (Bonferroni-corrected for multiple comparisons) revealed that the distributions of the combination of the log-ratio of neuron density and Euclidean distance ('density, distance') and the combination of the log-ratio of neuron density, Euclidean distance and the log-ratio of thickness ('density, distance, thickness') were not significantly different from each other ($p$ > .05), while the combination of the log-ratio of neuron density and Euclidean distance had a higher mean $J$ than all other combinations ($p$ < .05 for all pair-wise tests).

According to these results, we adopted the combination of the absolute log-ratio of neuron density and Euclidean distance as predictive variables for our probabilistic model. Figure 3A depicts the posterior probability for a projection to be present across the predictive variable space for this feature combination. Cross-validated classification performance across the evaluated thresholds is shown in the remainder of Figure 3. As shown in Figure 3B, classification accuracy quickly exceeded 80%, with a sizable fraction of the test set being classified. At higher thresholds, accuracy notably surpassed 90%, although this was accompanied by a decrease in the fraction of classified observations. As shown in Supplementary Figure S1, higher thresholds were associated with a consistent decrease in the rate of false positive predictions at an overall high rate of true positive predictions, resulting in a favorable Youden-index $J$ (Fig. 3C).

 Classification performance at all thresholds reliably exceeded chance performance as assessed by a permutation analysis. The permutation analysis revealed a classification performance from nonsensical labels that showed a relatively uniform accuracy of about 65% across tested thresholds. True positive rate and false positive rate equaled 1 across all thresholds, resulting in a Youden-index $J = 0.0 \pm 0.0$ for all thresholds.

Using the posterior probabilities obtained by the trained classifier (Fig 3A), we were able to make predictions about the status of projections between area pairs that were considered as unknown in the current data set [38]. We classified unknown projections at the threshold *p(threshold)* = 0.85, as indicated by the black lines in Figure 3A. Projections predicted to be absent or present are listed in Supplementary Table S1.

**Laminar patterns of projection neurons**

We observed a strong correlation between the fraction of labeled neurons originating in supragranular layers ($N_{SG}\%$) and log-ratio$_{density}$ ($r = 0.59$, $p < .001$, Fig 4A), as well as a moderate correlation between $N_{SG}\%$ and log-ratio$_{thickness}$ ($r = -0.42$, $p < .001$, Fig 4B). Given the strong correlation between the neuron density ratio and cortical thickness ratio, we computed a partial correlation of $N_{SG}\%$, log-ratio$_{density}$, and log-ratio$_{thickness}$ to assess the relative contribution of each variable. The partial correlation revealed that the correlation between thickness ratio and laminar patterns was mainly driven by the neuron density ratio, since the correlation did not reach significance when controlled for neuron density similarity ($r = 0.06$, $p > .05$). In contrast, the correlation between the neuron density ratio and laminar patterns was still significant when controlled for the cortical thickness ratio ($r = 0.43$, $p < .001$). Additionally, both $N_{SG}\%$ ($r = 0.09$, $p > .05$, Fig 4C) and $|N_{SG}\%|$ ($r = 0.003$, $p > .05$, Fig 4D) were independent of distance. Thus, the only anatomical factor that was systematically associated with laminar projection patterns was the architectonic similarity of linked areas.

**Relation of cytoarchitecture with connection topology**

We found that nodal network properties of cortical areas were related to the areas'

cytoarchitecture. Specifically, areas belonging to the structural network core had lower neuron density than non-core areas ($t_{(22)} = 2.9$, $p < .01$, $r = 0.52$, Fig 5A). Given that a major defining feature of core areas is their high degree (i.e., the large total number of connections), we tested whether this observation was indicative of a general relationship between cytoarchitectonic differentiation and the connectivity of areas. This analysis revealed that neuron density was strongly correlated with areal degree of connectivity ($r = -0.60$, $p < .01$, Fig 5B).

Additionally, we tested whether the same relationships could be observed for cortical thickness. Here the results were inconsistent. While cortical thickness did not differ between core and non-core areas ($t_{(27)} = -2.0$, $p > .05$, $r = 0.35$), thickness was moderately correlated with the degree of connectivity of areas ($r = 0.38$, $p < .05$).

Furthermore, we compared the neuron density and cortical thickness of five structural network modules that are related to spatial and functional sub-divisions of the cortex (specifically, comprising frontal, temporal, somato-motor, parieto-motor and occipito-temporal regions). These modules or clusters are characterized by denser structural connectivity within than between the modules [40]. Module assignments were taken from [41], who delineated the modules for a sub-network of 29x29 cortical areas [38] using a spectral decomposition algorithm. We found that the network modules differed in their neuron density ($H = 13.7$, $p < .01$), but not in their cortical thickness ($H = 7.2$, $p > .05$). *Post hoc* tests, Bonferroni-corrected for multiple comparisons, revealed that the frontal module had a higher neuron density than the occipito-temporal module ($t_{(13)} = 3.8$, $p = 0.002$, $r = 0.73$, $\alpha_{corr} = 0.005$); all other pairwise differences in neuron density between the modules were not significant after correcting for multiple comparisons.

**The architectonic basis of the primate connectome**

Architectonic differentiation defined by neuronal density of areas was the structural factor that related most consistently and strongly to the investigated features of the primate connectome. Figure 6 summarizes this finding and displays all present projections that were included in the analyses. Areas are arranged according to their neuron density, and projections are color-coded according to the neuron density ratio, expressing the architectonic similarity of the connected areas (from green for the smallest ratios via blue to purple for the highest density ratios). Note the dominance of projections linking architectonically similar areas (green links). Also note that core areas (indicated in red), are clustered at the lower end of the neuron density scale, as are areas with a relatively large number of connections, marked by their larger node size.

## Discussion

We assessed the extent to which distinct anatomical features can be used to predict the connectivity in the cerebral cortex of a non-human primate, using the most extensive quantitative data set of connections for the macaque monkey [38,39]. Specifically, we considered the cytoarchitectonic differentiation of cortical areas, quantified by neuron density, the spatial proximity of areas, quantified by Euclidean distance, and cortical thickness extracted from structural MRI data. We found that the existence of projections between areas depends strongly on the extent of their architectonic similarity (Fig 2A). We capitalized on this association to predict the existence of projections based on the structural relationships of potentially connected areas. Integrating cytoarchitectonic similarity and spatial proximity in a predictive model made it possible to determine whether two areas would be connected with more than 90% accuracy (Fig 3B). The model showed that a connection was most likely to exist between areas that are similar in their cytoarchitectonic differentiation and spatially close (Fig 3A). Our classification procedure consistently performed above chance level, as assessed by a permutation analysis. We used this classification procedure to make predictions about the status of unsampled projections (Supplementary Table S1), which provides an opportunity to compare our model's performance with future experimental results, allowing further model validation. Classification from alternative feature combinations revealed that when the three parameters were used as single predictors, cytoarchitectonic similarity yielded the highest maximum Youden-index *J* compared to Euclidean distance or thickness similarity on their own (Supplementary Figure S2B). This suggests that the performance of the model hinged predominantly on cytoarchitectonic similarity and to a lesser extent on spatial proximity. While thickness similarity also correlated with the relative frequency of present projections (Fig 2C), including this feature into our predictive model did not improve classification performance. Furthermore, even though the relative thickness of brain areas correlated strongly with the areas' relative neuron density, substituting density similarity for thickness similarity led to a considerable decrease in our model's predictive power.

Importantly, our model also revealed that, although the likelihood of a connection decreased across large differences in cytoarchitecture or long distances, this effect was mitigated if areas were spatially very close or respectively very similar in their cytoarchitecture. Thus, although connections were relatively less likely to exist between spatially remote areas, they did occur preferentially when distance was compensated for by similar cytoarchitectonic differentiation. Axonal wiring costs are a major constraint on structural connectivity [42] but are not strictly minimized in neural networks [6,43,44]. Our results highlight cytoarchitectonic differentiation as a key factor for predicting the occurrence of costly connections between spatially remote areas.

Additionally, we found that the laminar patterns of projection origins across the whole macaque cortex were very well explained by cytoarchitectonic similarity (Fig 4A), consistent with previous reports [8,9,19,20]. In contrast, there was no systematic relationship between laminar origin patterns of projections and distance or cortical thickness when the correlation with cytoarchitectonic differentiation was accounted for.

Moreover, we found that cytoarchitecture was closely associated with some of the essential topological properties of cortical areas. Specifically, areas belonging to the structural network core had a lower neuronal density than areas in the periphery (Fig 5A). This finding complements the observation that there are differences in several aspects of regional cellular morphology (e.g., dendritic tree size) between core and periphery areas [45]. One of the main defining features of core areas is their exceptionally large number of connections [46,47]. Therefore, we assessed whether there exists a direct relationship between cytoarchitecture as expressed by neuronal density and area degree (i.e., the number of connections of an area), without interposing the classification into core and periphery areas. This analysis revealed a strong general relationship between area degree and cytoarchitecture across the entire cortex. Thus, areas of lower density possessed a larger number of connections (Fig 5B), consistent with previous findings [13]. In contrast, cortical thickness showed an inconsistent and weaker relationship to membership in the structural network core and area degree.

It is not clear why there is a strong relationship between cortical cytoarchitecture and topological network features of areas, but answers are likely to be found in ontogeny. The development of the regional architectonic structure may be associated with the establishment of the connections of an area. One possible mechanism might draw on the relative timing of the emergence of areas, where areas that appear earlier might have the opportunity to connect more widely [14]. Indeed, a similar process has been suggested to explain the degree distribution of single neurons in *Caenorhabditis elegans* [48].

Additionally, we observed that network modules of areas differ in their cytoarchitecture. It has been suggested that network modules of the primate cortex result from a combination of spatial and topological properties [49]. Our findings suggest that cytoarchitecture may be another factor in the formation of structural modules, in line with our general conclusion that cortical architecture governs the formation of connections between brain areas.

While thickness measures have the advantage of being accessible non-invasively using MRI in humans, their relation to other anatomical features and to structural connectivity remains unclear. Our findings suggest that, while cortical thickness may show similarities to neuron density in its variability across the cerebral cortex, it is an imperfect surrogate and does not capture the fundamental aspects of brain networks that can be delineated from cytoarchitectonic differentiation.

In conclusion, our findings suggest that several features of the primate cortical connectome can largely be accounted for by the underlying structural properties of the cerebral cortex. Specifically, the relative cytoarchitectonic differentiation of the cortex provides an essential scaffold for explaining the organization of structural brain networks.

**General principles of cortical organization across mammalian species**

Does a model for explaining connections based on cytoarchitecture apply across species? The present results in the macaque cortex very closely parallel previous findings for the cat [20]. In both species, cytoarchitectonic differentiation was closely associated with multiple aspects of the organization of cortical networks, and cytoarchitectonic similarity integrated with spatial proximity was highly predictive of the existence of projections between potentially connected brain areas. This close association of brain architecture with connectivity was observed for areas distributed across the entire cortical surface, and was not contingent on grouping the areas into functional or anatomical modules of any kind. Moreover, cytoarchitectonic similarity has been consistently shown to explain laminar patterns of projections across macaque and cat [8,9,18–20,50]. Furthermore, an inverse relationship between the cytoarchitectonic differentiation and the connection degree of areas was observed in both species. Thus, areas of weaker differentiation have more connections. Highly connected areas are often hubs or members of a functionally prominent rich-club, occupying a topologically special position within networks of anatomical connections (e.g., van den Heuvel et al., 2012; van den Heuvel and Sporns, 2013b). Moreover, weakly differentiated areas likely differ from more strongly differentiated areas in their intrinsic circuitry and signal processing properties [51]. In combination, these findings indicate that the relative architectonic differentiation of cortical areas might shape the formation of corticocortical connections and thus impose constraints on structural as well as functional aspects of the connectome.

There is thus excellent correspondence of findings across two mammalian species and across the entire cerebral cortex. This evidence suggests that the reported association between architectonic differentiation of cortical areas and features of the inter-areal brain network reflects general organizational principles underlying the formation and maintenance of connections in the mammalian cortex.

**Conclusions**

Cytoarchitecture, which encompasses characteristic differences of local cortical organization [17], has previously been shown to explain laminar patterns of corticocortical connections [8,9,19,20]. Our results further underscore the significance of cytoarchitecture as a central factor that governs multiple aspects of the configuration of brain networks. This conclusion is based on three key observations about cortical connectivity: cortical cytoarchitecture is closely associated with the presence or absence of connections between cortices, the number of connections of a cortical area, as well as the laminar pattern of connections. By contrast, other factors such as cortical thickness and distance are not consistently related to connection features. The applicability of the structural model across different mammalian species and cortical systems suggests that it captures fundamental organizational principles underlying the global structural connectivity of the cerebral cortex. In humans, connections cannot be measured directly by tracing studies, but brain architecture can be studied *post mortem*. Thus, these findings also have important implications for understanding the structural connectivity of the human brain.

## Methods

We first introduce the analysed data of primate corticocortical connectivity and then present the structural parameters that were hypothesized to constrain connectivity. Subsequently, we describe measures and procedures used in the analyses.

### Connectivity data: Presence of projections

We used comprehensive data about corticocortical connectivity in the macaque brain obtained from systematic anatomical tracing experiments [38]. Briefly, the authors injected retrograde tracers into 29 cortical areas (parcellated according to their M132 atlas) and quantified labeled neurons found in all 91 areas of the M132 atlas that project to these injected sites. Within each area, labeled neurons ranged from a minimum of 1 neuron to a maximum of 262,279 neurons. Each of these is called a 'projection' to refer to a pathway from an area with labeled neurons to the injection site. The resulting data set contains information about the existence (i.e., either presence or absence) of 2610 projections within a 91x29 subgraph of the complete (91x91) connectivity matrix of the M132 atlas. For projections found to be present, projection strength is given as the fraction of labeled neurons, normalizing the number of projection neurons between two areas to the total number of labeled neurons for the respective injection, as done previously (e.g., [9,50]).

Crucially, the data set includes a 29x29 subgraph of injected areas, which contains information about all possible connections among the injected areas. This edge-complete subgraph makes it possible to perform analyses without uncertainty related to possible connections that were not sampled. Due to the wide distribution of the injected areas across the cortex, the 29x29 subgraph is expected to have similar properties as the complete network which incorporates all 91 areas [25]. Ercsey-Ravasz and colleagues [25] used the edge-complete subgraph to identify areas belonging to a 'network core' with a high density of connections among areas. This network core is similar to the concept of a rich-club, as discussed in recent studies [52–60]. Ercsey-Ravasz and colleagues [25] identified 17 core areas in the 29x29 subgraph, assigning the remaining 12 areas to the network periphery. We computed the degree of each area in the subgraph as the sum of the number of afferent and efferent projections of the area.

**Connectivity data: Laminar origin of projection neurons**

In addition, we analyzed the laminar patterns of projection origins in 11 areas [39], which Markov and colleagues extracted from the set of 29 injections described above [38]. Here, the fraction of labeled neurons originating in supragranular layers ($N_{SG}\%$) was provided for 625 projections originating in 11 of the 29 injected areas and targeting all 91 areas of the M132 atlas. Specifically, $N_{SG}\%$ was computed as the number of supragranular labeled neurons divided by the sum of supragranular and infragranular labeled neurons [39]. To relate $N_{SG}\%$ to the undirected measure of Euclidean distance, we also transformed it to an undirected measure of inequality in laminar patterns, $|N_{SG}\%|$, where $|N_{SG}\%| = |N_{SG}\%-50|*2$. Values of $N_{SG}\%$ around 0% and 100% thus translated to larger values of $|N_{SG}\%|$, indicating a more pronounced inequality in the distribution of origins of projection neurons between infra- and supragranular layers. We based our analyses regarding $N_{SG}\%$ on the subset of 429 projections comprising more than 20 neurons (neuron numbers for each projection are provided in [38]). Thus, we excluded very weak projections for which assessment of the distribution of projection neurons in cortical layers was not considered reliable (cf. [18]). Results did not change qualitatively if a less conservative threshold of 10 neurons was applied.

**Structural model: Neuron density**

The spectrum of architectonic differentiation ranges from areas of low overall neuron density, with few layers and lacking an inner granular layer (agranular), to dense areas with six distinct layers. The striate cortex, for example, has a much higher overall neuron density not only within the cortical visual system, but also among all other cerebral cortices [11,61–66]. Intermediate to these two extremes are areas of lower neuron densities with a sparse inner granular layer (dysgranular), and areas with six layers but without the exceptional clarity of layers and sublayers or remarkable neuron density of striate cortex. We used an unbiased quantitative stereologic approach to study the cytoarchitecture of each area expressed by neuron density. We estimated neuron density from coronal sections of macaque cortex that were stained to mark neurons using either Nissl stain or immunohistochemical staining for neuronal nuclei-specific antibody (NeuN), which labels neurons but not glia, using a microscope-computer interface (StereoInvestigator, MicroBrightField Inc., Williston, VT). We verified that there is a close correspondence between measures derived from both staining methods in a sample of areas for which both measures were available ($r = 0.99$, $p = 0.001$), and accordingly transformed density measures from different staining methods to a common reference frame. The neuron density measurements used here have partly been published previously [14]. In total, neuron density measures were available for 48 areas (Fig 1). Within the 29x29 subgraph, neuron densities were available for 14 of the 17 core areas and 10 of the 12 non-core areas. We quantified relative cytoarchitectonic differentiation across the cortex by neuron density. We computed the log-ratio of neuron density values for each pair of connected areas (which is equivalent to the difference of the logarithms of the area densities), where log-ratio$_{density}$ = log (density$_{source\ area}$ / density$_{target\ area}$). The use of a logarithmic scale was indicated, since the most extreme value of the neuron density measures was more than three standard deviations above the mean of the considered neuron densities [67]. For analyses which required considering an undirected equivalent of the actual neuron density ratio, we used the absolute value of the log-ratio, |log-ratio$_{density}$|. From the available neuron density measures we were able to determine the relative cytoarchitectonic profile for 1128 of the sampled projections, including 172 projections with an associated $N_{SG}$%.

**Distance model: Spatial proximity**

We operationalized the spatial proximity of all 91 cortical areas by the Euclidean distance between their mass centers, obtained from the Scalable Brain Atlas (http://scalablebrainatlas.incf.org). This widely used interval measure of projection length represents a pragmatic estimate of the spatial proximity of pre- and postsynaptic neurons located in different brain areas (e.g., [68–74]. Information about the spatial proximity of areas was included for all 2610 sampled projections, also encompassing all 429 projections we analyzed with respect to |$N_{SG}$%|.

**Thickness model: cortical thickness**

Cortical thickness data were extracted from an anatomical T1-weighted magnetic resonance (MR) brain scan of one male adult macaque monkey (*Macaca mulatta*). MR data were acquired on a 3 Tesla Philips Achieva MRI scanner using a three-dimensional magnetization prepared rapid acquisition gradient-echo (3DMPRAGE) sequence with 0.6 mm isotropic voxels (130 slices, TR = 7.09 ms, TE = 3.16 ms, FOV = 155 x 155 mm²). Cortical reconstruction and volumetric segmentation were performed using the Freesurfer image analysis suite (http://surfer.nmr.mgh.harvard.edu/). The resulting surface reconstruction was registered to the M132 atlas [38] using the Caret software [75] (http://www.nitrc.org/projects/caret/). Cortical thickness was then extracted for all 91 areas in both hemispheres using Freesurfer. Here, we report results for mean thickness values of the left and right hemisphere.

The thickness measurements extracted from MR data were well correlated with microscopic measurements of histological sections [14]. Corresponding histological and MR measurements for 33 areas were available, resulting in $r = 0.62$, $p < 0.001$ for the left hemisphere, $r = 0.48$, $p < 0.01$ for the right hemisphere, and $r = 0.56$, $p < 0.001$ for mean thickness values of the left and right hemisphere.

To quantify relative thickness across the cortex in order to compare thickness in pairs of connected areas, we computed the log-ratio of thickness values for each pair of areas analogous to the log-ratio of neuron density, where
log-ratio$_{thickness}$ = log (thickness$_{source\ area}$ / thickness$_{target\ area}$). We transformed the log-ratio of cortical thickness to an undirected equivalent, |log-ratio$_{thickness}$|, where appropriate. Relative thickness of areas was included for all 2610 sampled projections, also encompassing all 429 projections analyzed with respect to $N_{SG}$%.

**Relative projection frequencies**

To characterize the distribution of present and absent projections across the range of each anatomical variable, while accounting for differences in sampling, we computed relative frequencies of projections that were present. Particularly, we partitioned each anatomical variable into bins and normalized the number of present projections in each bin by the total number of studied projections (i.e., absent and present projections that fall into the respective bin). This procedure allowed us to obtain a measure of the relative frequency of present projections which is robust against disparities in sampling across a variable's range (e.g., when more projections were sampled across a spatial separation of 10 – 15 mm than across 50 – 55 mm, as can be seen from the absolute projection numbers). We verified that results were robust against changes in bin size.

**Classification of projection existence**

We combined the anatomical variables in a probabilistic predictive model for classifying the existence of projections. We built this model using a binary support vector machine (SVM) classifier (i.e., used for two-class learning), which received the anatomical variables associated with the projections as independent variables (features) and information about projection existence (i.e., projection status 'absent' or 'present') as the dependent variable (labels, comprising two classes). Euclidean distance, absolute log-ratio of neuron density and absolute log-ratio of cortical thickness were used as features in different combinations.

For training the SVM classifier, we used a linear kernel function, standardized the independent variables prior to classification and assumed uniform prior probabilities for the learned classes. Classification scores obtained from the trained classifier were transformed to the posterior probability that an observation was classified as 'present', *p(present)*. To assess performance of the classification procedure, we used five-fold cross-validation. We randomly partitioned all available observations into five folds of equal size. After training the SVM classifier on a training set comprising four folds, we used the resulting posterior probabilities to predict the status of the remaining fold (20% of available observations) that comprised the test set. We used two classification rules derived from a common threshold probability. (1) We assigned the status 'present' to all observations whose posterior probability exceeded the threshold probability, that is, observations with *p(present) > p(threshold)*. (2) We assigned the status 'absent' to all observations with *p(present) < 1 - p(threshold)*. The approach was applied to thresholds from *p(threshold)* = 0.50 to *p(threshold)* = 1.00, in increments of 0.025. By increasing the threshold probability, we therefore narrowed the windows in the feature space for which classification was possible. For thresholds of *p(threshold) <= 0.50*, the classification windows overlap. In particular, parts of the feature space corresponding to classification as 'present' overlap with parts corresponding to classification as 'absent', and observations would therefore be classified twice. For this reason, we did not consider thresholds below *p(threshold)* = 0.50. For each threshold, we computed performance as described below and averaged results across the five cross-validation folds. To make performance assessment robust against variability in the partitioning of observations, we report performance measures averaged across 100 rounds of the five-fold cross-validation.

We assessed classification performance by computing prediction accuracy, the fraction of correct predictions relative to all predictions. Accuracy was also separately assessed for positive and negative predictions, yielding precision and negative predictive value as the fraction of correct positive or correct negative predictions relative to all positive or negative predictions, respectively. We also computed which fraction of observations in the test set was assigned a prediction at a given threshold. As further performance measures, we computed sensitivity (true positive rate) and specificity (true negative rate) at the evaluated thresholds. We also computed the false positive rate (1-specificity). To quantify performance based on sensitivity and specificity, we computed the Youden-index $J$ as $J$ = sensitivity + specificity – 1 [76,77]. $J$ is a measure of how well a binary classifier operates above chance level, with $J = 0$ indicating chance performance and $J = 1$ indicating perfect classification. Since $J$ is defined at each threshold, to obtain a single summary measure we computed the mean of $J$ across the more conservative thresholds *p(present)* = 0.85 to *p(present)* = 1.00 for all 100 cross-validation runs. Results did not change if the maximum $J$ across all thresholds was considered instead (Supplementary Fig. S2B).

To assess statistical null performance of the classification procedure, we performed a permutation analysis. The analysis was equal to the classification procedure described above, with the exception of an additional step prior to the partitioning of observations into cross-validation folds. Here, for each round of cross-validation, the labels were randomly permuted. Thereby, the correspondence between features and true labels of observations was removed. In the permutation analysis, we used Euclidean distance and the absolute log-ratio of neuron density as features, based on the feature combination that led to the best results, and averaged performance measures across 1000 rounds of five-fold cross-validation.

These analyses were performed using Matlab R2014a (The MathWorks, Inc., Natick, MA,

**Statistical tests**

To test groups of projections for equality in their associated anatomical variables, we computed two-tailed independent samples t-tests and report the t-statistic *t*, degrees of freedom *df* and the associated measure of effect size *r*, where $r = (t^2/(t^2+df))^{1/2}$. Results did not change if Welch's t-test was applied, which does not assume equal variances across groups. To test for equality of more than two groups of areas regarding their neuron density or cortical thickness, we computed the non-parametric Kruskal–Wallis test statistics (*H*). To assess relations between interval variables, we computed Pearson's correlation coefficient *r*. For ordinal variables, we computed Spearman's rank-correlation coefficient ρ. All tests were pre-assigned a two-tailed significance level α = 0.05. These analyses were performed using Matlab R2012b (The MathWorks, Inc., Natick, MA, United States).

**Methodological considerations**

Some comments need to be made on the anatomical variables used in the present analysis. First, we used overall neuron density of brain areas to capture the complex architectonic profile of different cortices in a single parameter. Other crucial features of cytoarchitecture include the number and distinctiveness of cortical layers and the relative width and granularity of layer 4. Additionally, features that cannot be observed in cytoarchitecture, for example myeloarchitectonic properties, contribute to a fuller characterization of cortical differentiation (see [78]. However, many of these aspects are difficult to quantify. Moreover, there exists no consistent framework for integrating these measures into a one-dimensional ranking of structural differentiation. In practice, estimates of the overall differentiation of brain areas rely on subjective expert categorizations (resulting in the assignment of areas to 'structural types' (cf. [9,20]. By contrast, neuron density can be determined objectively using unbiased stereologic methods. In a comparison of multiple quantitative features of cortical architecture, neuron density turned out to be the most discriminating parameter for identifying cortical areas in the primate prefrontal cortex [14]. The features included in the analysis comprised cortical thickness, and density of different cell markers, including neurons, glia, and neurons labeled with calbindin, calretinin or parvalbumin, and their respective laminar distributions. Further, there is a close correspondence between neuron density measurements and expert ratings of cytoarchitectonic differentiation that comprehensively take into account multiple dimensions [14]. Thus, neuron density is a well established, characteristic measure for quantifying cytoarchitectonic differentiation in sensory and high-order association cortices.

Second, we used measurements of cortical thickness obtained from structural MRI in one macaque monkey. The MRI measures provided coverage of all cortical areas, and agreed well with the corresponding microscopic thickness measurements from histological sections (cf. Methods section Thickness model: cortical thickness). This finding is in line with similar agreements between histological and MRI-based thickness measures seen for cortical regions of the human brain [79]. Therefore, the thickness measurements were considered reliable, despite the small sample size. Reliability was further strengthened by averaging thickness values for corresponding regions of the left and right hemisphere. We found that the correlation between the distribution of projection origins and relative cortical thickness was mainly driven by differences in neuron density, as elaborated in Results and Discussion.

# References

1. Park, H.-J. & Friston, K. J. Structural and Functional Brain Networks: From Connections to Cognition. *Science* **342,** 1238411 (2013).
2. Rockland, K. S. & Pandya, D. N. Laminar origins and terminations of cortical

connections of the occipital lobe in the rhesus monkey. *Brain Res.* **179,** 3–20 (1979).

3. Pandya, D. N. & Yeterian, E. H. in *Association and auditory cortices* (eds. Peters, A. & Jones, E. G.) **4,** 3–61 (Plenum Press, 1985).

4. Felleman, D. J. & Van Essen, D. C. Distributed hierarchical processing in the primate cerebral cortex. *Cereb. Cortex* **1,** 1–47 (1991).

5. Hilgetag, C. C., O'Neill, M. A. & Young, M. P. Indeterminate organization of the visual system. *Science* **271,** 776–777 (1996).

6. Bullmore, E. T. & Sporns, O. Complex brain networks: graph theoretical analysis of structural and functional systems. *Nat. Rev. Neurosci.* **10,** 186–198 (2009).

7. Colizza, V., Flammini, A., Serrano, M. A. & Vespignani, A. Detecting rich-club ordering in complex networks. *Nat. Phys.* **2,** 110–115 (2006).

8. Barbas, H. Pattern in the laminar origin of corticocortical connections. *J. Comp. Neurol.* **252,** 415–422 (1986).

9. Barbas, H. & Rempel-Clower, N. L. Cortical structure predicts the pattern of corticocortical connections. *Cereb. Cortex* **7,** 635–646 (1997).

10. Sanides, F. in *The Primate Brain* (eds. Noback, C. R. & Montagna, W.) 137–208 (Appleton-Century-Crofts, 1970).

11. Pandya, D. N., Seltzer, B. & Barbas, H. in *Comparative Primate Biology, Vol.4: Neurosciences* (eds. Steklis, H. & Erwin, J.) 39–80 (Alan R. Liss, 1988).

12. Barbas, H. General Cortical and Special Prefrontal Connections: Principles from Structure to Function. *Annu. Rev. Neurosci.* **38,** 269–289 (2015).

13. Barbas, H. & Pandya, D. N. Architecture and intrinsic connections of the prefrontal cortex in the rhesus monkey. *J. Comp. Neurol.* **286,** 353–375 (1989).

14. Dombrowski, S. M., Hilgetag, C. C. & Barbas, H. Quantitative Architecture Distinguishes Prefrontal Cortical Systems in the Rhesus Monkey. *Cereb. Cortex* **11,** 975–988 (2001).

15. von Economo, C. F. *Zellaufbau der Grosshirnrinde des Menschen*. (Springer, 1927).

16. von Economo, C. *Cellular Structure of the Human Cerebral Cortex*. (Karger Medical and Scientific Publishers, 2009).

17. Zilles, K. & Amunts, K. Segregation and Wiring in the Brain. *Science* **335,** 1582–1584 (2012).

18. Barbas, H., Hilgetag, C. C., Saha, S., Dermon, C. R. & Suski, J. L. Parallel organization of contralateral and ipsilateral prefrontal cortical projections in the rhesus monkey. *BMC Neurosci.* **6,** 32 (2005).

19. Hilgetag, C. C. & Grant, S. Cytoarchitectural differences are a key determinant of laminar projection origins in the visual cortex. *NeuroImage* **51,** 1006–1017 (2010).

20. Beul, S. F., Grant, S. & Hilgetag, C. C. A predictive model of the cat cortical connectome based on cytoarchitecture and distance. *Brain Struct. Funct.* **220,** 3167–3184 (2015).

21. Young, M. P. Objective analysis of the topological organization of the primate cortical visual system. *Nature* **358,** 152–155 (1992).

22. Klyachko, V. A. & Stevens, C. F. Connectivity optimization and the positioning of cortical areas. *Proc. Natl. Acad. Sci.* **100,** 7937–7941 (2003).

23. Markov, N. T. *et al.* The role of long-range connections on the specificity of the macaque interareal cortical network. *Proc. Natl. Acad. Sci.* **110,** 5187–5192 (2013).

24. Douglas, R. J. & Martin, K. A. C. Mapping the Matrix: The Ways of Neocortex. *Neuron* **56,** 226–238 (2007).

25. Ercsey-Ravasz, M. *et al.* A Predictive Network Model of Cerebral Cortical Connectivity Based on a Distance Rule. *Neuron* **80,** 184–197 (2013).

26. Salin, P. A. & Bullier, J. Corticocortical connections in the visual system: structure and function. *Physiol. Rev.* **75,** 107–154 (1995).

27. Cullen, T. J. *et al.* Anomalies of asymmetry of pyramidal cell density and structure in

dorsolateral prefrontal cortex in schizophrenia. *Br. J. Psychiatry* **188,** 26–31 (2006).

28. la Fougère, C. *et al.* Where in-vivo imaging meets cytoarchitectonics: The relationship between cortical thickness and neuronal density measured with high-resolution [18F]flumazenil-PET. *NeuroImage* **56,** 951–960 (2011).

29. Narr, K. L. *et al.* Mapping Cortical Thickness and Gray Matter Concentration in First Episode Schizophrenia. *Cereb. Cortex* **15,** 708–719 (2005).

30. Lerch, J. P. *et al.* Mapping anatomical correlations across cerebral cortex (MACACC) using cortical thickness from MRI. *NeuroImage* **31,** 993–1003 (2006).

31. He, Y., Chen, Z. J. & Evans, A. C. Small-World Anatomical Networks in the Human Brain Revealed by Cortical Thickness from MRI. *Cereb. Cortex* **17,** 2407–2419 (2007).

32. Gong, G., He, Y., Chen, Z. J. & Evans, A. C. Convergence and divergence of thickness correlations with diffusion connections across the human cerebral cortex. *NeuroImage* **59,** 1239–1248 (2012).

33. Chen, Z. J., He, Y., Rosa-Neto, P., Germann, J. & Evans, A. C. Revealing Modular Architecture of Human Brain Structural Networks by Using Cortical Thickness from MRI. *Cereb. Cortex* **18,** 2374–2381 (2008).

34. Chen, Z. J., He, Y., Rosa-Neto, P., Gong, G. & Evans, A. C. Age-related alterations in the modular organization of structural cortical network by using cortical thickness from MRI. *NeuroImage* **56,** 235–245 (2011).

35. Bernhardt, B. C., Klimecki, O. M., Leiberg, S. & Singer, T. Structural Covariance Networks of the Dorsal Anterior Insula Predict Females' Individual Differences in Empathic Responding. *Cereb. Cortex* **24,** 2189–2198 (2014).

36. Tewarie, P. *et al.* Disruption of structural and functional networks in long-standing multiple sclerosis. *Hum. Brain Mapp.* (2014). doi:10.1002/hbm.22596

37. Evans, A. C. Networks of anatomical covariance. *NeuroImage* **80,** 489–504 (2013).

38. Markov, N. T. *et al.* A Weighted and Directed Interareal Connectivity Matrix for

Macaque Cerebral Cortex. *Cereb. Cortex* **24,** 17–36 (2014).

39. Markov, N. T. *et al.* The anatomy of hierarchy: Feedforward and feedback pathways in macaque visual cortex: Cortical counter-streams. *J. Comp. Neurol.* **522,** 225–259 (2014).

40. Hilgetag, C. C., Burns, G. A. P. C., O'Neill, M. A., Scannell, J. W. & Young, M. P. Anatomical connectivity defines the organization of clusters of cortical areas in the macaque monkey and the cat. *Philos. Trans. R. Soc. B Biol. Sci.* **355,** 91–110 (2000).

41. Goulas, A., Schaefer, A. & Margulies, D. S. The strength of weak connections in the macaque cortico-cortical network. *Brain Struct. Funct.* **220,** 2939–2951 (2014).

42. Bullmore, E. T. & Sporns, O. The economy of brain network organization. *Nat. Rev. Neurosci.* **13,** 336–349 (2012).

43. Chen, B. L., Hall, D. H. & Chklovskii, D. B. Wiring optimization can relate neuronal structure and function. *Proc. Natl. Acad. Sci. U. S. A.* **103,** 4723–4728 (2006).

44. Kaiser, M. & Hilgetag, C. C. Nonoptimal Component Placement, but Short Processing Paths, due to Long-Distance Projections in Neural Systems. *PLoS Comput. Biol.* **2,** e95 (2006).

45. Scholtens, L. H., Schmidt, R., de Reus, M. A. & Van Den Heuvel, M. P. Linking Macroscale Graph Analytical Organization to Microscale Neuroarchitectonics in the Macaque Connectome. *J. Neurosci.* **34,** 12192–12205 (2014).

46. Hagmann, P. *et al.* Mapping the Structural Core of Human Cerebral Cortex. *PLoS Biol* **6,** e159 (2008).

47. Harriger, L., van den Heuvel, M. P. & Sporns, O. Rich Club Organization of Macaque Cerebral Cortex and Its Role in Network Communication. *PLoS ONE* **7,** e46497 (2012).

48. Varier, S. & Kaiser, M. Neural Development Features: Spatio-Temporal Development of the Caenorhabditis elegans Neuronal Network. *PLoS Comput. Biol.* **7,** e1001044 (2011).

49. Chen, Y., Wang, S., Hilgetag, C. C. & Zhou, C. Trade-off between Multiple Constraints Enables Simultaneous Formation of Modules and Hubs in Neural Systems. *PLoS Comput

*Biol* **9,** e1002937 (2013).

50. Medalla, M. & Barbas, H. Diversity of laminar connections linking periarcuate and lateral intraparietal areas depends on cortical structure. *Eur. J. Neurosci.* **23,** 161–179 (2006).

51. Beul, S. F. & Hilgetag, C. C. Towards a 'canonical' agranular cortical microcircuit. *Front. Neuroanat.* **8,** 165 (2015).

52. van den Heuvel, M. P., Kahn, R. S., Goñi, J. & Sporns, O. High-cost, high-capacity backbone for global brain communication. *Proc. Natl. Acad. Sci.* **109,** 11372–11377 (2012).

53. Crossley, N. A. *et al.* Cognitive relevance of the community structure of the human brain functional coactivation network. *Proc. Natl. Acad. Sci.* **110,** 11583–11588 (2013).

54. Tomasi, D., Wang, R., Wang, G.-J. & Volkow, N. D. Functional Connectivity and Brain Activation: A Synergistic Approach. *Cereb. Cortex* bht119 (2013). doi:10.1093/cercor/bht119

55. Towlson, E. K., Vértes, P. E., Ahnert, S. E., Schafer, W. R. & Bullmore, E. T. The Rich Club of the C. elegans Neuronal Connectome. *J. Neurosci.* **33,** 6380–6387 (2013).

56. van den Heuvel, M. P. & Sporns, O. Network hubs in the human brain. *Trends Cogn. Sci.* **17,** 683–696 (2013).

57. van den Heuvel, M. P. & Sporns, O. An Anatomical Substrate for Integration among Functional Networks in Human Cortex. *J. Neurosci.* **33,** 14489–14500 (2013).

58. Ball, G. *et al.* Rich-club organization of the newborn human brain. *Proc. Natl. Acad. Sci.* **111,** 7456–7461 (2014).

59. Collin, G., Sporns, O., Mandl, R. C. W. & van den Heuvel, M. P. Structural and Functional Aspects Relating to Cost and Benefit of Rich Club Organization in the Human. *Cereb. Cortex* **24,** 2258–2267 (2014).

60. Crossley, N. A. *et al.* The hubs of the human connectome are generally implicated in the

anatomy of brain disorders. *Brain J. Neurol.* **137,** 2382–2395 (2014).

61. Zilles, K. in *Neurobiologie psychischer Störungen* (eds. Förstl, P. D. med H., Hautzinger, P. D. D.-P. M. & Roth, P. D. phil D. rer nat G.) 75–140 (Springer Berlin Heidelberg, 2006).

62. O'Kusky, J. & Colonnier, M. A laminar analysis of the number of neurons, glia, and synapses in the adult cortex (area 17) of adult macaque monkeys. *J. Comp. Neurol.* **210,** 278–290 (1982).

63. Schüz, A. & Palm, G. Density of neurons and synapses in the cerebral cortex of the mouse. *J. Comp. Neurol.* **286,** 442–455 (1989).

64. Collins, C. E., Airey, D. C., Young, N. A., Leitch, D. B. & Kaas, J. H. Neuron densities vary across and within cortical areas in primates. *Proc. Natl. Acad. Sci.* **107,** 15927–15932 (2010).

65. Cahalane, D. J., Charvet, C. J. & Finlay, B. L. Systematic, balancing gradients in neuron density and number across the primate isocortex. *Front. Neuroanat.* **6,** 28 (2012).

66. Herculano-Houzel, S., Watson, C. & Paxinos, G. Distribution of neurons in functional areas of the mouse cerebral cortex reveals quantitatively different cortical zones. *Front. Neuroanat.* **7,** 35 (2013).

67. Buzsáki, G. & Mizuseki, K. The log-dynamic brain: how skewed distributions affect network operations. *Nat. Rev. Neurosci.* **15,** 264–278 (2014).

68. Salvador, R. *et al.* Neurophysiological Architecture of Functional Magnetic Resonance Images of Human Brain. *Cereb. Cortex* **15,** 1332–1342 (2005).

69. Achard, S., Salvador, R., Whitcher, B., Suckling, J. & Bullmore, E. T. A Resilient, Low-Frequency, Small-World Human Brain Functional Network with Highly Connected Association Cortical Hubs. *J. Neurosci.* **26,** 63–72 (2006).

70. Bassett, D. S. *et al.* Hierarchical Organization of Human Cortical Networks in Health and Schizophrenia. *J. Neurosci.* **28,** 9239–9248 (2008).

71. Alexander-Bloch, A. F. *et al.* The Anatomical Distance of Functional Connections Predicts Brain Network Topology in Health and Schizophrenia. *Cereb. Cortex* **23,** 127–138 (2013).

72. Goñi, J. *et al.* Resting-brain functional connectivity predicted by analytic measures of network communication. *Proc. Natl. Acad. Sci.* **111,** 833–838 (2014).

73. Vértes, P. E. *et al.* Simple models of human brain functional networks. *Proc. Natl. Acad. Sci.* **109,** 5868–5873 (2012).

74. Tewarie, P. *et al.* Structural degree predicts functional network connectivity: A multimodal resting-state fMRI and MEG study. *NeuroImage* **97,** 296–307 (2014).

75. Van Essen, D. C. *et al.* An Integrated Software Suite for Surface-based Analyses of Cerebral Cortex. *J. Am. Med. Inform. Assoc.* **8,** 443–459 (2001).

76. Youden, W. J. Index for rating diagnostic tests. *Cancer* **3,** 32–35 (1950).

77. Fluss, R., Faraggi, D. & Reiser, B. Estimation of the Youden Index and its Associated Cutoff Point. *Biom. J.* **47,** 458–472 (2005).

78. Barbas, H. & García-Cabezas, M. Á. Motor cortex layer 4: less is more. *Trends Neurosci.* **38,** 259–261 (2015).

79. Scholtens, L. H., de Reus, M. A. & van den Heuvel, M. P. Linking contemporary high resolution magnetic resonance imaging to the von economo legacy: A study on the comparison of MRI cortical thickness and histological measurements of cortical structure. *Hum. Brain Mapp.* **36,** 3038–3046 (2015).

## Acknowledgments

We thank Alexandros Goulas for helpful suggestions on the analyses and Konrad Wagstyl for the extraction of cortical thickness values from MR data. SB and CCH were supported by DFG Collaborative Research Center Grant SFB 936/A1. HB is supported by grants from NIH (NINDS R01NS024760; and NIMH, R01MH057414) and CELEST, an NSF Science of Learning Center (NSF OMA-0835976). The authors declare no competing financial interests.

## Author Contributions

SB and CCH designed the study. SB conducted neural density measures and performed the analyses. HB provided neural density data. All authors discussed the results and wrote the manuscript.

## Competing interests

The authors declare no competing financial interests.

Figures

**Figure 1**: Neuron densities in the macaque cortex depicted on the M132 parcellation (Markov et al., 2014a). Grey areas: no density data available. Abbreviations as in Markov et al. (2014a).

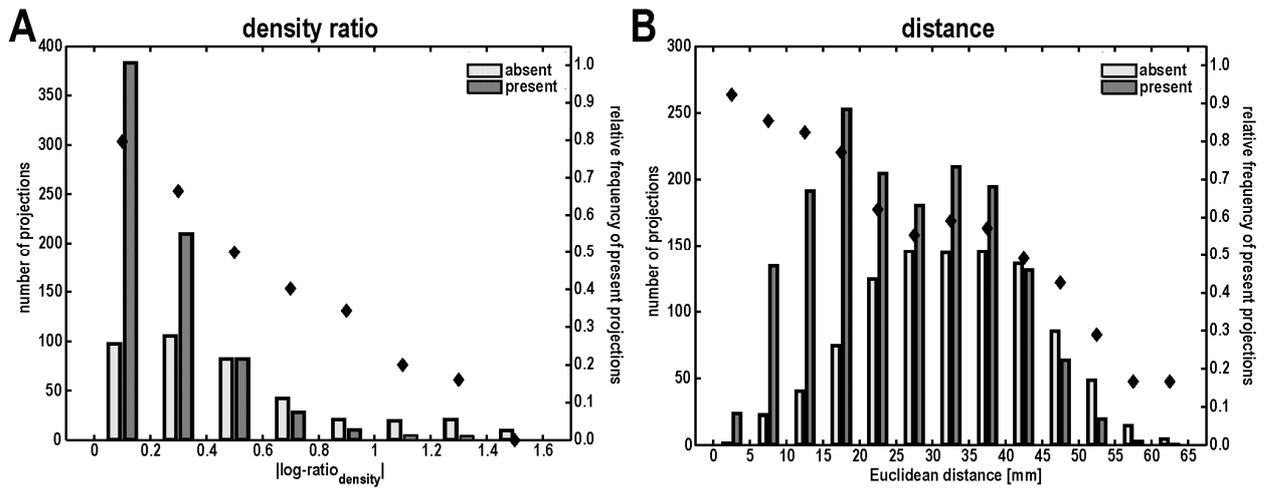

**Figure 2**: Distribution of absent and present projections across neuron density ratio (A) and Euclidean distance (B). Absolute numbers of absent and present projections (bars) are depicted alongside the corresponding relative frequency of present projections (diamonds).

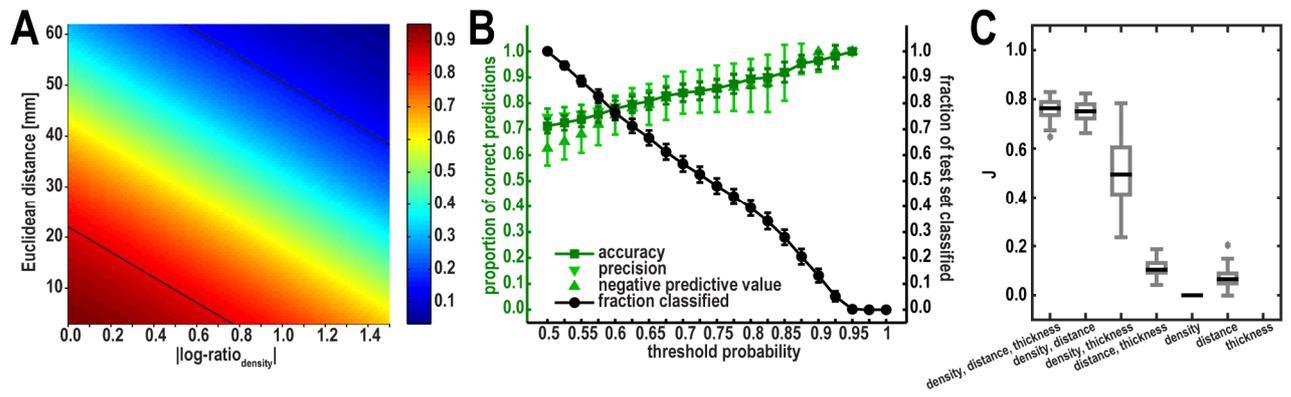

**Figure 3**: (A,B) Classification of projection existence from |log-ratio$_{density}$| and Euclidean distance. (A) Posterior probability of a projection being present resulting from training the classifier on all projections. Black lines are positioned at *p(present)* = 0.85 and *p(present)* = 0.15. Also see Supplementary Table S1 for predictions made about unsampled projections at these thresholds. (B) Cross-validated classification performance at different thresholds. Mean prediction accuracy for projections that were predicted to be present and absent (light green) as well as overall mean prediction accuracy (dark green) are shown. Also shown is the fraction of the test set that was classified at each threshold (black). Error bars indicate standard deviations. (C) Youden-index *J* for all combinations of parameters. Distribution of mean *J* across thresholds *p(present)* = 0.85 to *p(present)* = 1.00 for all 100 rounds of cross-validation. Boxplots indicate median *J* by a black bar and outliers by gray circles.

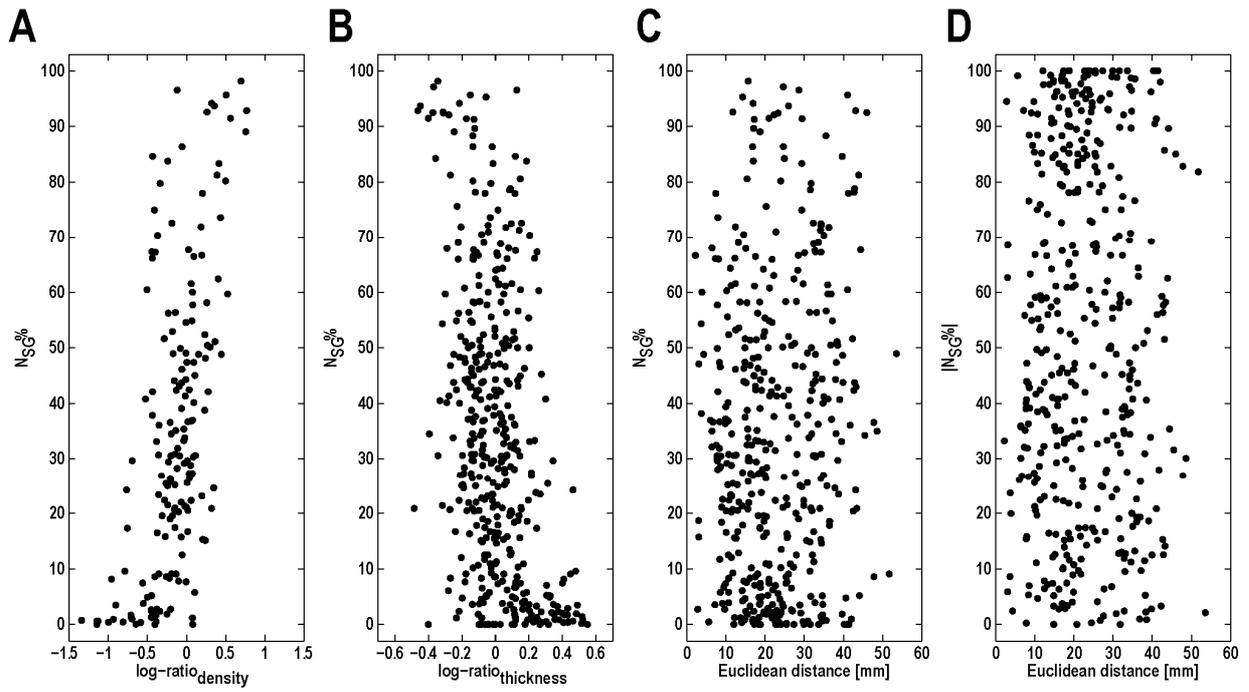

**Figure 4**: Variation of laminar patterns of projection origins with anatomical variables. The fraction of labeled projection neurons originating from supragranular layers $N_{SG}\%$ was strongly correlated with log-ratio$_{density}$ (A) and moderately correlated with log-ratio$_{thickness}$ (B). Neither $N_{SG}\%$ nor $|N_{SG}\%|$ was correlated with Euclidean distance (C, D).

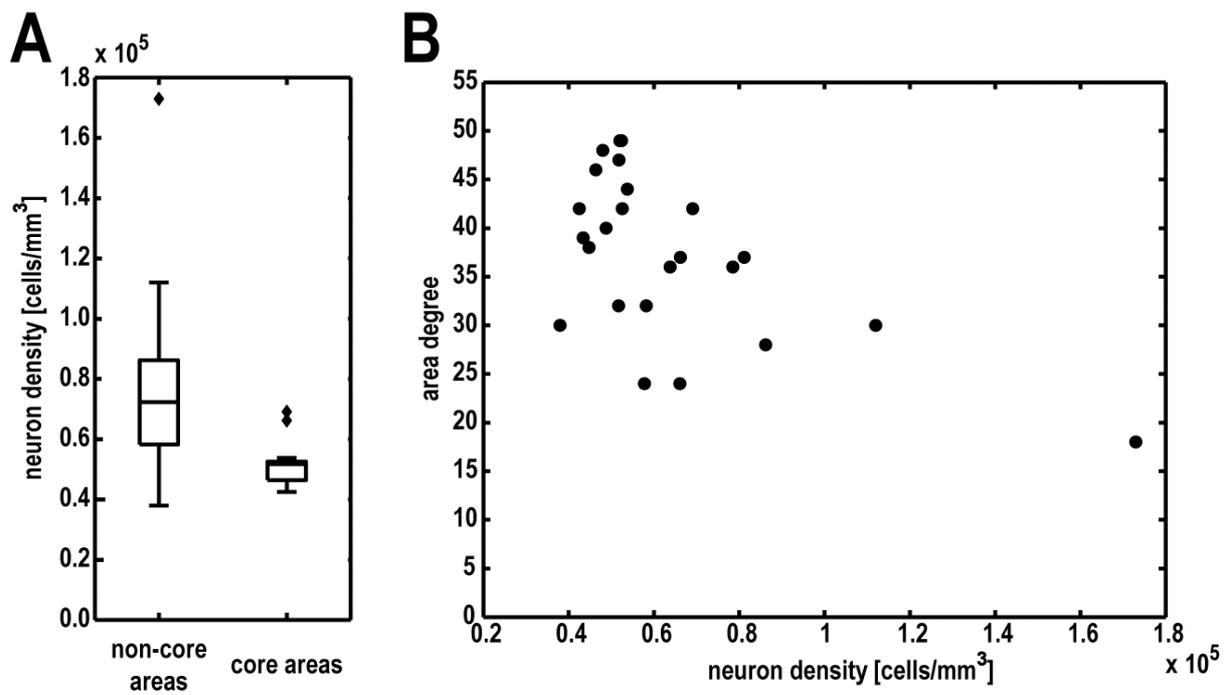

**Figure 5**: Variation of topological properties with neuron density. (A) Areas that were identified as belonging to a structural core network by Ercsey-Ravasz and colleagues (2013) had a significantly lower neuron density than non-core areas. (B) The number of connections maintained by an area (area degree) decreases with increasing neuron density.

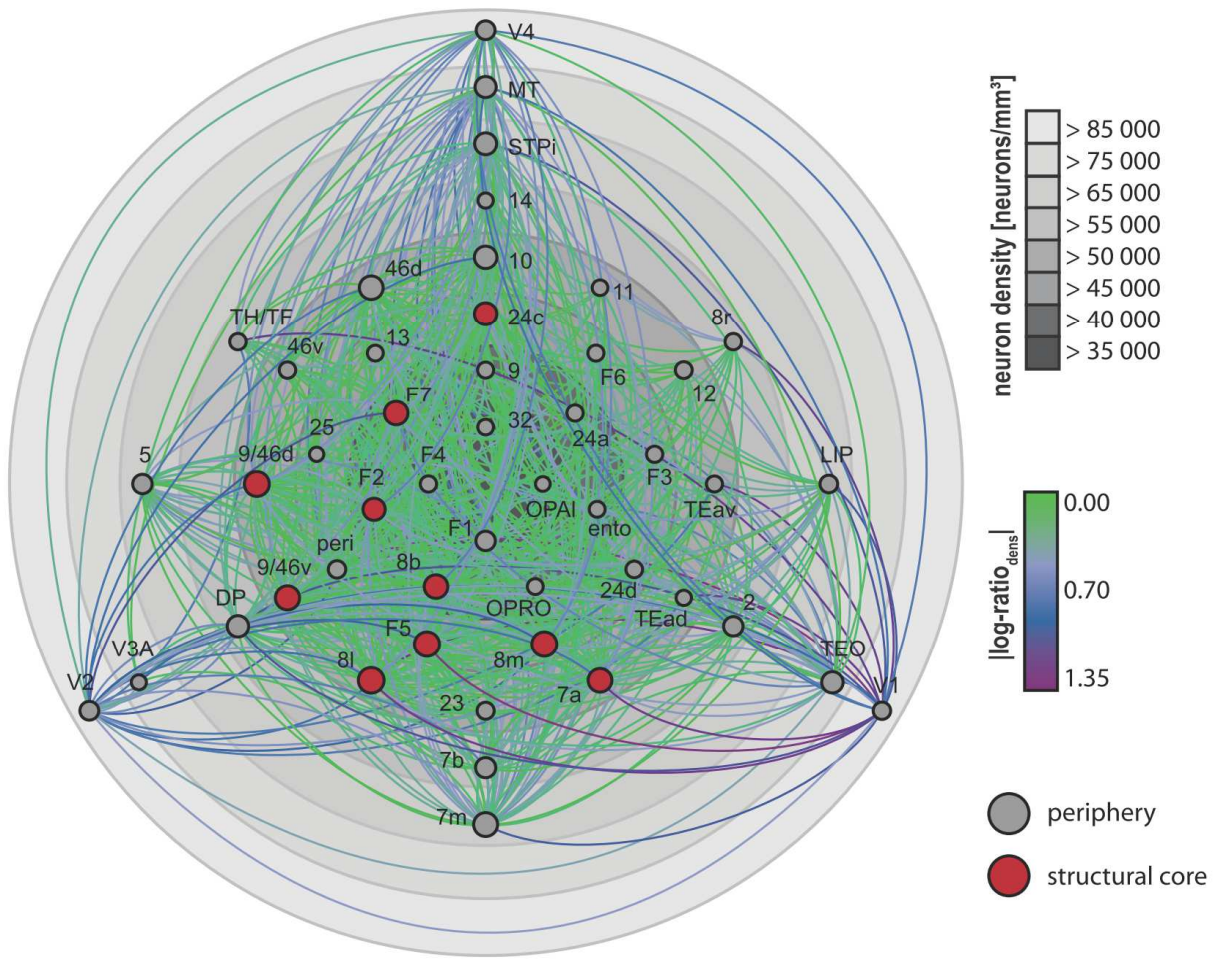

**Figure 6**: Primate cortical connectome visualized based on neuron density gradients. Grey circles correspond to neuron density, increasing from center to periphery; cortical areas are positioned accordingly (cf. Fig. 1). Present projections between cortical areas are displayed color-coded according to absolute neuron density ratios of the connected areas from green (small ratios) via blue to purple (large ratios). Node sizes indicate the areas' degree. Structural core areas, as classified by Ercsey-Ravasz et al. (2013), are filled in red. Abbreviations as in Markov et al. (2014a).

# A Predictive Structural Model of the Primate Connectome

### Sarah F. Beul[1], Helen Barbas[2,3], Claus C. Hilgetag[1,2]

[1] Department of Computational Neuroscience, University Medical Center Hamburg-Eppendorf, Martinistr.52 – W36, 20246 Hamburg, Germany

[2] Neural Systems Laboratory, Department of Health Sciences, Boston University, Commonwealth Ave. 635, 20115 Boston, MA, USA

[3] Boston University School of Medicine, Department of Anatomy and Neurobiology, 72 East Concord St., 02118 Boston, MA, USA

## Supplementary Data

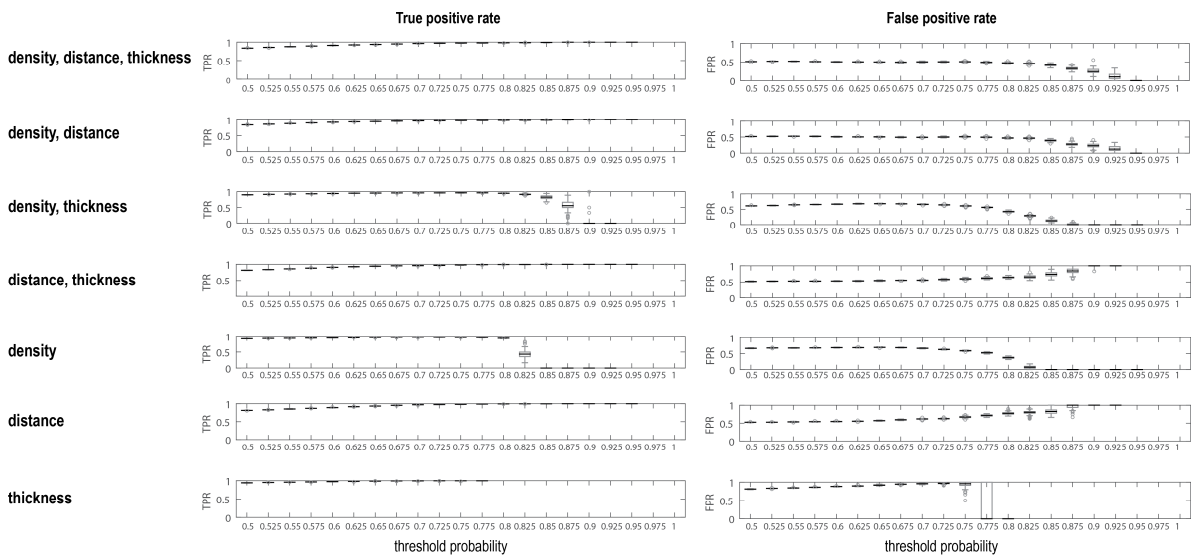

**Supplementary Figure S1:** True positive rate and false positive rate for classification of projection existence from all possible combinations of parameters. Distribution of rates across all 100 rounds of cross-validation is shown for all threshold probabilities. Overall performance was best for the combination of |log-ratio$_{density}$| and Euclidean distance. Note that the addition of |log-ratio$_{thickness}$| to these two parameters did not improve performance. Boxplots indicate median rates by a black bar and outliers by gray circles.

Beul, Barbas, Hilgetag (2016) A Predictive Structural Model of the Primate Connectome

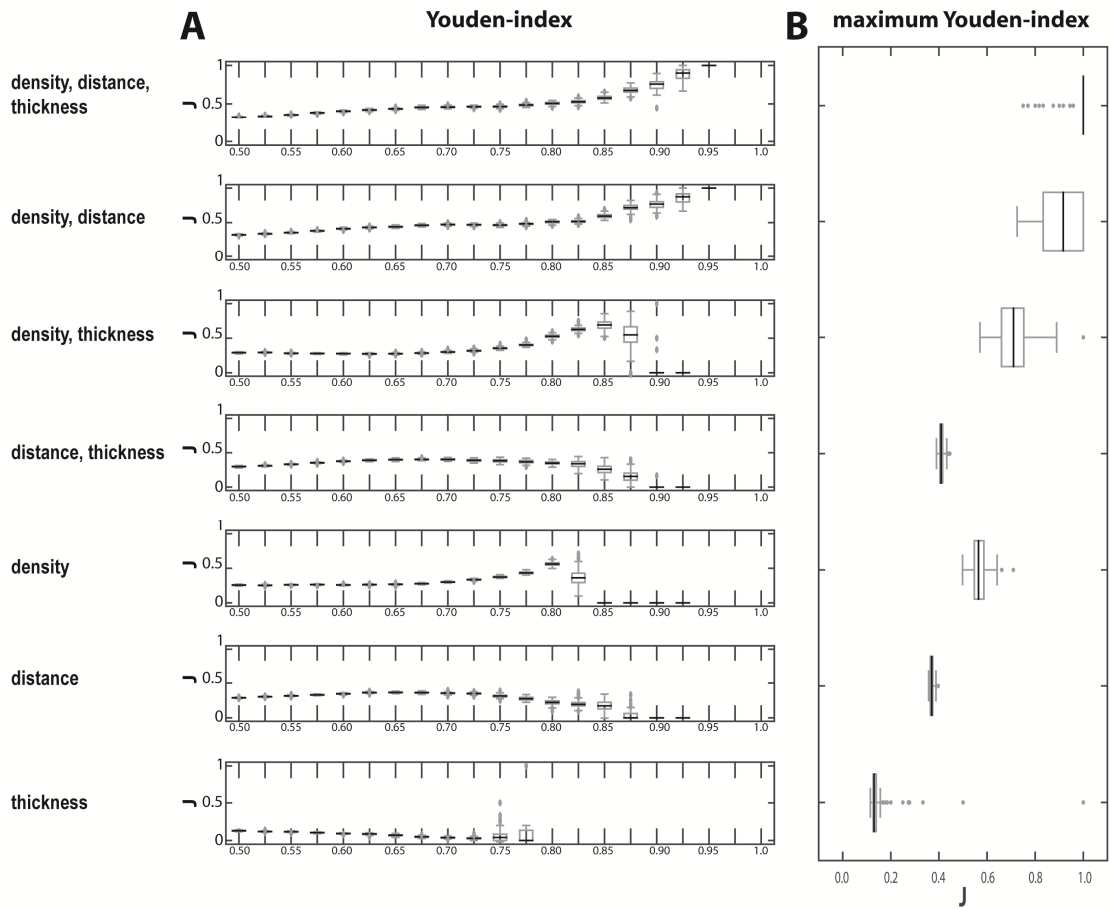

**Supplementary Figure S2:** Youden-index *J* for classification of projection existence from all possible combinations of parameters. (A) Distribution of *J* across all 100 rounds of cross-validation is shown for all threshold probabilities. Overall performance was best for the combination of |log-ratio$_{density}$| and Euclidean distance. Note that the addition of |log-ratio$_{thickness}$| to these two parameters did not improve performance. (B) Distribution of maximum *J* (across all threshold probabilities) for all 100 rounds of cross-validation. Kruskal-Wallis-test showed that the distributions were significantly different ($H = 661.0$, $p < .001$). *Post hoc* tests (Bonferroni-corrected) revealed that the distributions of 'density, distance, thickness' and 'density, distance' were not significantly different from each other ($p > .05$), while all other pair-wise tests reached statistical significance (all *p* < .05). Boxplots indicate median *J* by a black bar and outliers by gray circles.

Beul, Barbas, Hilgetag (2016) A Predictive Structural Model of the Primate Connectome

**Supplementary Table S1, related to Figure 3**: Classification of unsampled projections. The status of projections not sampled in the data set was predicted from the posterior probabilities resulting from the trained classifier (Fig 3). Projections were predicted to be absent if their associated |log-ratio$_{density}$| and Euclidean distance yielded a posterior probability for a projection to be present of *p(present)* <= 0.15, and predicted to be present if *p(present)* >= 0.85.

| Projections predicted to be absent | | Source area | Target area | Source area | Target area | Source area | Target area | Source area | Target area |
|---|---|---|---|---|---|---|---|---|---|
| | | 9/46v | 9 | F6 | 12 | F6 | 14 | 9/46d | 32 |
| Source area | Target area | F2 | 9 | F7 | 12 | OPAI | 14 | F3 | 32 |
| V1 | 9 | F3 | 9 | OPAI | 12 | 24d | 23 | F4 | 32 |
| V2 | 9 | F6 | 9 | OPRO | 12 | 8b | 23 | F6 | 32 |
| V1 | 11 | F7 | 9 | 9 | 13 | 8l | 23 | F7 | 32 |
| V1 | 12 | 9 | 11 | 10 | 13 | 8m | 23 | OPAI | 32 |
| V1 | 13 | 10 | 11 | 11 | 13 | F1 | 23 | OPRO | 32 |
| V1 | 14 | 12 | 11 | 12 | 13 | F2 | 23 | 9 | 24a |
| V1 | 25 | 13 | 11 | 14 | 13 | F3 | 23 | 11 | 24a |
| V1 | 32 | 14 | 11 | 25 | 13 | F4 | 23 | 12 | 24a |
| V2 | 32 | 25 | 11 | 32 | 13 | 9 | 25 | 13 | 24a |
| V1 | 24a | 32 | 11 | 24a | 13 | 10 | 25 | 14 | 24a |
| V1 | 24d | 24a | 11 | 24c | 13 | 11 | 25 | 25 | 24a |
| V1 | 46v | 24c | 11 | 24d | 13 | 12 | 25 | 32 | 24a |
| V1 | 8r | 24d | 11 | 46d | 13 | 13 | 25 | 24c | 24a |
| V1 | F3 | 46d | 11 | 46v | 13 | 14 | 25 | 24d | 24a |
| V1 | F4 | 46v | 11 | 8b | 13 | 32 | 25 | 46d | 24a |
| V1 | F6 | 8b | 11 | 8l | 13 | 24a | 25 | 46v | 24a |
| V1 | OPAI | 8l | 11 | 8m | 13 | 24c | 25 | 8b | 24a |
| V1 | OPRO | 8m | 11 | 8r | 13 | 24d | 25 | 8m | 24a |
| | | 8r | 11 | 9/46d | 13 | 46d | 25 | 9/46d | 24a |
| Projections predicted to be present | | 9/46d | 11 | 9/46v | 13 | 46v | 25 | F2 | 24a |
| | | 9/46v | 11 | F3 | 13 | 8b | 25 | F3 | 24a |
| Source area | Target area | F6 | 11 | F4 | 13 | 8m | 25 | F4 | 24a |
| 10 | 9 | F7 | 11 | F5 | 13 | 9/46d | 25 | F6 | 24a |
| 11 | 9 | 9 | 12 | F6 | 13 | 9/46v | 25 | F7 | 24a |
| 12 | 9 | 10 | 12 | F7 | 13 | F3 | 25 | OPAI | 24a |
| 13 | 9 | 11 | 12 | OPAI | 13 | F6 | 25 | OPRO | 24a |
| 14 | 9 | 13 | 12 | OPRO | 13 | OPAI | 25 | 9 | 24d |
| 25 | 9 | 14 | 12 | 9 | 14 | OPRO | 25 | 11 | 24d |
| 32 | 9 | 25 | 12 | 10 | 14 | 9 | 32 | 12 | 24d |
| 24a | 9 | 32 | 12 | 11 | 14 | 10 | 32 | 13 | 24d |
| 24c | 9 | 24a | 12 | 12 | 14 | 11 | 32 | 14 | 24d |
| 24d | 9 | 24c | 12 | 13 | 14 | 12 | 32 | 23 | 24d |
| 46d | 9 | 24d | 12 | 25 | 14 | 13 | 32 | 25 | 24d |
| 46v | 9 | 46d | 12 | 32 | 14 | 14 | 32 | 32 | 24d |
| 8b | 9 | 46v | 12 | 24a | 14 | 25 | 32 | 24a | 24d |
| 8m | 9 | 8b | 12 | 24c | 14 | 24a | 32 | 24c | 24d |
| 8r | 9 | 8l | 12 | 24d | 14 | 24c | 32 | 46d | 24d |
| 9/46d | 9 | 8m | 12 | 46d | 14 | 24d | 32 | 46v | 24d |
| | | 8r | 12 | 46v | 14 | 46d | 32 | 8b | 24d |
| | | 9/46d | 12 | 8r | 14 | 46v | 32 | 8l | 24d |
| | | 9/46v | 12 | 9/46d | 14 | 8b | 32 | 8m | 24d |
| | | F5 | 12 | 9/46v | 14 | 8m | 32 | 8r | 24d |

Beul, Barbas, Hilgetag (2016) A Predictive Structural Model of the Primate Connectome

| Source area | Target area | Source area | Target area | Source area | Target area | Source area | Target area |
|---|---|---|---|---|---|---|---|
| 9/46d | 24d | peri | ento | 24a | F6 | TEO | TEad |
| 9/46v | 24d | TEad | ento | 24c | F6 | TH/TF | TEad |
| F2 | 24d | TEav | ento | 24d | F6 | 2 | TEav |
| F3 | 24d | TH/TF | ento | 46d | F6 | ento | TEav |
| F4 | 24d | 9 | F3 | 46v | F6 | peri | TEav |
| F6 | 24d | 13 | F3 | 8b | F6 | TEad | TEav |
| F7 | 24d | 23 | F3 | 8l | F6 | teo | TEav |
| 9 | 46v | 25 | F3 | 8m | F6 | TH/TF | TEav |
| 10 | 46v | 32 | F3 | 8r | F6 | ento | TH/TF |
| 11 | 46v | 24a | F3 | 9/46d | F6 | peri | TH/TF |
| 12 | 46v | 24c | F3 | 9/46v | F6 | TEad | TH/TF |
| 13 | 46v | 24d | F3 | F2 | F6 | TEav | TH/TF |
| 14 | 46v | 46d | F3 | F3 | F6 | 7m | V3a |
| 25 | 46v | 46v | F3 | F4 | F6 | DP | V3a |
| 32 | 46v | 8b | F3 | F7 | F6 | LIP | V3a |
| 24a | 46v | 8l | F3 | 5 | LIP | MT | V3a |
| 24c | 46v | 8m | F3 | 7a | LIP | V2 | V3a |
| 24d | 46v | 8r | F3 | 7m | LIP | V4 | V3a |
| 46d | 46v | 9/46d | F3 | DP | LIP | | |
| 8b | 46v | F1 | F3 | STPi | LIP | | |
| 8l | 46v | F2 | F3 | V3a | LIP | | |
| 8m | 46v | F4 | F3 | 12 | OPAl | | |
| 8r | 46v | F6 | F3 | 13 | OPAl | | |
| 9/46d | 46v | F7 | F3 | 14 | OPAl | | |
| 9/46v | 46v | 13 | F4 | 25 | OPAl | | |
| F3 | 46v | 23 | F4 | 32 | OPAl | | |
| F6 | 46v | 32 | F4 | 24a | OPAl | | |
| F7 | 46v | 24a | F4 | 8b | OPAl | | |
| 2 | 8r | 24d | F4 | F4 | OPAl | | |
| 9 | 8r | 8b | F4 | OPRO | OPAl | | |
| 11 | 8r | 8l | F4 | 12 | OPRO | | |
| 12 | 8r | 8m | F4 | 13 | OPRO | | |
| 13 | 8r | 8r | F4 | 25 | OPRO | | |
| 14 | 8r | 9/46d | F4 | 32 | OPRO | | |
| 24c | 8r | F1 | F4 | 24a | OPRO | | |
| 24d | 8r | F2 | F4 | ento | OPRO | | |
| 46d | 8r | F3 | F4 | F4 | OPRO | | |
| 46v | 8r | F5 | F4 | F5 | OPRO | | |
| 8b | 8r | F6 | F4 | OPAl | OPRO | | |
| 8l | 8r | F7 | F4 | peri | OPRO | | |
| 8m | 8r | OPAl | F4 | 2 | peri | | |
| 9/46d | 8r | OPRO | F4 | ento | peri | | |
| 9/46v | 8r | 9 | F6 | F5 | peri | | |
| F2 | 8r | 10 | F6 | OPRO | peri | | |
| F3 | 8r | 11 | F6 | TEad | peri | | |
| F4 | 8r | 12 | F6 | TEav | peri | | |
| F5 | 8r | 13 | F6 | TH/TF | peri | | |
| F6 | 8r | 14 | F6 | ento | TEad | | |
| F7 | 8r | 25 | F6 | peri | TEad | | |
| OPRO | ento | 32 | F6 | TEav | TEad | | |

4